\documentclass[a4paper,12pt]{article}
\usepackage{graphicx}
\global\arraycolsep=2pt 
\input{epsf}
\begin{document}
\makeatletter
\def\fmslash{\@ifnextchar[{\fmsl@sh}{\fmsl@sh[0mu]}}
\def\fmsl@sh[#1]#2{%
  \mathchoice
    {\@fmsl@sh\displaystyle{#1}{#2}}%
    {\@fmsl@sh\textstyle{#1}{#2}}%
    {\@fmsl@sh\scriptstyle{#1}{#2}}%
    {\@fmsl@sh\scriptscriptstyle{#1}{#2}}}
\def\@fmsl@sh#1#2#3{\m@th\ooalign{$\hfil#1\mkern#2/\hfil$\crcr$#1#3$}}
\newcommand{\beq}{\begin{equation}}
\newcommand{\eeq}{\end{equation}}
\newcommand{\beqa}{\begin{eqnarray}}
\newcommand{\eeqa}{\end{eqnarray}}
\def\r){\right)}
\def\l({\left(}
\def\nn{\nonumber}
\makeatother
\thispagestyle{empty}
\begin{titlepage}

\begin{flushright}
hep-ph/0104278 \\
LMU 01/08 \\
\today
\end{flushright}

\vspace{0.3cm}
\boldmath
\begin{center}
  { \Large \bf Kaluza-Klein Theories and the Anomalous Magnetic Moment
    of the Muon}
\end{center}
\unboldmath
\vspace{0.8cm}
\begin{center}
 {\large Xavier Calmet
   \footnote{e-mail:calmet@theorie.physik.uni-muenchen.de}, Andrey Neronov
   \footnote{e-mail:neronov@theorie.physik.uni-muenchen.de}}\\
  \end{center}
\begin{center}
{\sl Ludwig-Maximilians-University Munich, Sektion Physik}\\
{\sl Theresienstra{\ss}e 37, D-80333 Munich, Germany}\\
\end{center}

\vspace{1cm}

\begin{abstract}
\noindent 
We discuss nonminimal couplings of fermions to the electromagnetic
field, which generically appears in models with extra dimensions. We
consider models where the electromagnetic field is generated by the
Kaluza-Klein mechanism. The nonminimal couplings contribute at
tree-level to anomalous magnetic and electric dipole moments of
fermions. We use recent measurements of these quantities to put limits
on the parameters of models with extra dimensions.
\end{abstract}

\end{titlepage}

The measurements of the electric and anomalous magnetic dipole moments
of fermions provide a very stringent test of the Standard Model of
particle physics. The latest measurement of $g-2$ of the muon
\cite{Brown:2001mg} seems to indicate a deviation from the Standard
Model. The difference between the experimental value and the
theoretical value calculated in the framework of the Standard Model
is:
\begin{eqnarray} \label{expres}
  \Delta{a_\mu}=a_\mu(\mbox{exp})-a_\mu(\mbox{SM}) &=& 43 (16) \times 10^{-10}.
\end{eqnarray}
For a theoretical review see \cite{Czarnecki:2001pv}. This effect does
not necessarily imply a conflict with the Standard Model as there are
theoretical uncertainties in the calculations involving hadronic
quantum corrections \cite{Yndurain:2001qw}.  Interpretations of this
$2.6\ \sigma$ effect were immediately proposed in different frameworks
like compositeness and technicolor \cite{comp}, supersymmetry
\cite{susy}, extra dimensions \cite{extra}, massive neutrinos
\cite{mneutrino}, leptoquarks \cite{lq}, additional gauge bosons
\cite{egboson} and others \cite{other}. In this note we want to
discuss the contribution to the electric and to the anomalous magnetic
dipole moments of fermions which arises due to the nonminimal coupling
\beq
\label{anom1}
{\cal S}_{int}= \int d^4x
F_{\mu\nu}\overline\psi(A+B\gamma^5)\sigma^{\mu\nu} \psi \eeq of
fermions to the electromagnetic field $F_{\mu\nu}=\partial_\mu
A_\nu-\partial_\nu A_\mu$ ( $\sigma_{\mu \nu}=\frac{1}{4}[ \gamma_\mu
,\gamma_\nu ])$.  This nonminimal coupling generically appears
in Kaluza-Klein type theories \cite{nonmin}, e.g. Kaluza-Klein
supergravity \cite{sugra}.

In different  models with small, large and infinite extra dimensions
\cite{sugra,add,rs} the space-time manifold is taken to be a product of 
the four-dimensional Minkowski space 
$M^4$ and a Riemannian manifold $K^n$. In a coordinate chart 
$(x^\mu, y^\alpha)$ on $M^4\times K^n$, a general Lorentz-invariant 
metric can be written in the form
\beq
\label{background}
ds^2=f(y)\eta_{\mu\nu} dx^\mu dx^\nu+g_{\alpha\beta}(y)dy^\alpha
dy^\beta,
\eeq
where $\eta_{\mu\nu}$ is the four-dimensional Minkowski
metric and $g_{\alpha\beta}$ is a metric on $K^n$.  Depending on the
type of model under consideration, the observable matter fields of the
Standard Model of particle physics are either allowed to propagate in
the bulk, like in the conventional Kaluza-Klein theory and in certain
compactification schemes of higher-dimensional supergravity, or
confined to live on a four-dimensional surface ${\cal M}^4\subset
M^4\times K^n$. In the former case the typical size $L$ of compact
manifold $K^n$ is restricted to be small enough, so that the energy
scale of Kaluza-Klein excitations of Standard Model particles $M\sim
L^{-1}$ is higher than the experimentally reachable one, $M\geq 1$
TeV. In the latter case the size of $K^n$ can be large and even
infinite.

If the manifold $K^n$ possesses a nontrivial isometry group ${\cal G}$, 
perturbations of the background metric (\ref{background}) which have the form 
\begin{equation}
\label{kk}
g_{AB}=\left(
\begin{array}{cl}
f(y)\eta_{\mu\nu}+g_{\alpha\beta} A_\mu^\alpha A_\nu^\beta & \vline \ 
g_{\alpha\beta} A_\mu^\beta\\
\hline
g_{\alpha\beta}A_\mu^\beta &\vline\  g_{\alpha\beta}
\end{array}
\right)
\end{equation} 
where the vector fields $A^\alpha_\mu$  
\beq
\label{killing}
A_\mu^\alpha={\cal A}_\mu^{(p)}\xi_{(p)}^\alpha
\eeq
are proportional to  Killing vector fields 
$\xi^\alpha_{(p)}, \ p=1,..,$ dim ${\cal G}$. One
can, in principle, interpret these fields as observable gauge fields
of the Standard Model if the group ${\cal G}$ contains $SU(3)\times
SU(2)\times U(1)$ as a subgroup.  In this case the minimal dimension
of the compact manifold $K^n$ is $n=7$ \cite{witten}. Seven
dimensional compact manifolds with such isometry group were
extensively studied and classified in the context of compactifications
of eleven-dimensional supergravity \cite{sugra}.

In general, in models with large or infinite extra dimensions, when all the
fields of the Standard Model are supposed to be localized on a
four-dimensional brane, the fields
$A_\mu^\alpha$ cannot be related to the Standard Model gauge fields since
they
are generally not confined to the brane ${\cal M}^4$. But, it turns out that
a certain
subset of these fields, which is  associated to the symmetry of
rotations around the brane can be naturally localized on the brane
in models with warped extra dimensions (that is, when $f(y)\not=
1$) \cite{neronov}. Thus, with some modifications,  
Kaluza-Klein's idea of relating gauge fields to isometries of
higher-dimensional space-time can also be implemented in this type of
models.

If the observable electromagnetic field originates from isometries of
higher-dimensional space-time, it differs from the electromagnetic
field of Maxwell theory in a crucial point.  Although this field is
minimally coupled to scalars, the coupling to spin-1/2 fields contains
nonminimal terms (\ref{anom1}), \cite{nonmin}. This coupling results
in an additional contribution to the anomalous magnetic moment (if
$A\not= 0$) and to the electric (if $B\not= 0$) dipole moment of
fermions. The term proportional to $\gamma_5$ is $T$ and $P$
violating.

In the simplest (unrealistic) five-dimensional Kaluza-Klein theory the
parameters $A$ and $B$ are determined by the five-dimensional
gravitational constant $G_5$ and the size $L$ of the only extra
dimension.  $G_5$ and $L$ are, in turn, expressed through the
four-dimensional gravitational constant $G_4\sim 10^{-38}$ GeV$^{-2}$
and four-dimensional fine structure constant $\alpha_{QED} \sim
10^{-2}$. An order-of-magnitude estimate shows that the contribution
of the nonminimal coupling (\ref{anom1}) to the anomalous magnetic moment
and to the electric dipole moment is, in this case, too small to be
experimentally detected.
 
In higher dimensional theories the parameter space is richer and, as
we will show below, the strength of the anomalous coupling
(\ref{anom1}) depends not only on the geometry of $K^n$ and the
$4+n$-dimensional gravitational constant $G_{4+n}$ but also on the
form of the fermion zero modes on $K^n$ which give rise to the
observable four-dimensional fermions. In this case, the available
experimental data on anomalous magnetic and electric dipole moments
can provide certain restrictions on parameters of higher-dimensional
theory.

Couplings of Kaluza-Klein gauge fields to the fermion fields can be found 
from the higher-dimensional Dirac action   
\beq
\label{dirac}
{\cal S}_{D}=i  \int d^4xd^ny(\det E^{\hat M}_N)\ \overline\Psi
\Gamma^A D_A\Psi
\eeq
where $E^{\hat M}_N$ is the vielbein field (hat denotes the local Lorentz 
indexes and the bulk metric is expressed through the vielbein as 
$g_{AB}=E^{\hat C}_A E^{\hat D}_B\eta_{\hat C\hat D}$ where 
$\eta_{\hat C\hat D}$ is $4+n$-dimensional Minkowski metric). The curved 
space gamma matrices $\Gamma^A$ are related to the 
flat space ones as $\Gamma^A=E^A_{\hat B}\Gamma^{\hat B}$ where 
$\{\Gamma^{\hat B},\Gamma^{\hat C}\}=2\eta^{\hat B\hat C}$. The covariant 
derivative of Dirac spinor is defined as
\beq
D_A\Psi=\partial_A\Psi+\frac{1}{2}\omega^{\hat B\hat C}_A
\sigma_{\hat B\hat C}\Psi,
\eeq
where $\omega_A^{\hat B\hat C}$ is the spin connection expressed in a 
standard way through derivatives of the vielbein and 
$\sigma_{\hat B\hat C}=\frac{1}{4}[\Gamma_{\hat B}, \Gamma_{\hat C}]$ is 
generator of local Lorentz rotations. 

The $U(1)$ gauge group of electromagnetism is a one-parametric subgroup 
of  the isometry group ${\cal G}$ generated by a Killing vector
field $\xi^\alpha_U(y)$. 
Taking a coordinate $y^1=\theta$ along the integral curves of  
$\xi^\alpha_U(y)$ we can write the metric (\ref{kk}) as 
\begin{eqnarray}
\label{kk1}
g_{AB}=\left(
\begin{array}{cll}
f(y^a) \eta_{\mu\nu}+\varphi A_\mu A_\nu& \vline \ 
\varphi A_\mu &\vline\ 0\\
\hline 
\varphi A_\mu &\vline\  \varphi &\vline\ 0\\
\hline 0 &\vline\ 0 &\vline g_{ab}
\end{array}
\right)
\end{eqnarray} 
where $\varphi=\varphi(y^a), a=2,.., n$, since the metric coefficients do not 
depend on $\theta$.

Taking the coordinate vielbein for the metric (\ref{kk1}) we have
\beqa
\label{sc}
\Gamma^A D_A\Psi&=&\frac{1}{\sqrt{f}}\Gamma^{\hat \mu}
(\partial_\mu-A_\mu\partial_\theta)\Psi
+\frac{1}{\sqrt{\varphi}}\Gamma^{\hat\theta}\partial_\theta \Psi \nn\\
&& + \Gamma^a\l(D_a+\frac{f_{,a}}{f}+\frac{\varphi_{,a}}{4\varphi}\r)\Psi
+\frac{\sqrt{\varphi}}{8f}F_{\mu\nu}\Gamma^{\hat\mu}\Gamma^{\hat\nu}
\Gamma^{\hat\theta}\Psi
\eeqa
where $D_a$ is the covariant derivative with respect to the metric $g_{ab}$.
We take $\Psi$ of the form
\beq
\Psi=e^{iq\theta}\psi(x)\chi(y^a)
\eeq
where $q$ is an integer. $\psi$ is a four-dimensional spinor which 
satisfies the massless Dirac 
equation
\beq
i\Gamma^{\hat \mu}\partial_\mu\psi=0
\eeq
and  $\chi$ is a $n$-dimensional spinor which obeys
\beq
i\Gamma^a\l(D_a+\frac{f_{,a}}{f}+\frac{\varphi_{,a}}{4\varphi}\r)\chi
-\frac{q}{\sqrt{\varphi}}\Gamma^{\hat\theta}\chi =0
\eeq
For such configurations the  action (\ref{dirac}) reduces to
\beqa
\label{act}  \! \! \! \!  \! \! \! \!
{\cal S}= \int d^ny f^2\sqrt{\varphi|g_{ab}|}|\chi|^2
\int d^4x 
\overline\psi
\left[ \frac{i}{\sqrt{f}}\Gamma^{\hat \mu} 
(\partial _\mu-iqA_\mu)+\frac{i\sqrt{\varphi}}{4f}F_{\mu\nu}
\sigma^{\hat \mu\hat \nu}\Gamma^{\hat \theta}\right]\psi  
\eeqa
We can take $\Gamma^{\hat \mu}=\gamma^\mu, \Gamma^{\hat \theta}=\gamma^5$ where
$\gamma^\mu, \gamma^5$ are conventional four-dimensional gamma matrices.
The $4+n$-dimensional spinor $\Psi$ must be normalized with respect to the
norm
\beq
\left<\Psi,\Psi\right>=\int d^4xd^ny \sqrt{-g}\ \overline\Psi\Gamma^0\Psi,
\eeq
which yields  the following normalization condition for the
$n$-dimensional spinor $\chi$ 
\beq
\int d^ny f^{3/2}\sqrt{\varphi|g_{ab}|}|\chi|^2=1.
\eeq
From (\ref{act}) we see that the nonminimal coupling of the fermion field to 
$A_\mu$ has the form
\beq
\label{red_act}
{\cal S}_{int}= R_\chi \int d^4x
\overline\psi F_{\mu\nu}\sigma^{\mu\nu}\gamma^5\psi
\eeq
with
\beq
\label{R}
R_\chi=\frac{1}{4} \int d^ny f \varphi\sqrt{|g_{ab}|}|\chi|^2.
\eeq
Note, that the strength of the nonminimal coupling depends on the
higher-dimensional profile $\chi$ of the fermion zero mode and, therefore,
it can be different for different fermions. The interpretation
of $R_\chi$ depends on the particular model under consideration. For example,
in models with large or infinite extra
dimensions and localized Kaluza-Klein gauge fields \cite{neronov},
$R_\chi$ characterises the size of the  region where
the fermion $\Psi$ is localized.

In the absence of $F_{\mu \nu}$, the action (\ref{act}) is invariant under
chiral rotations 
\beq
\label{chirrot}
P\psi=e^{-i\alpha\gamma^5}\psi \eeq where $\alpha$ is an arbitrary
angle. Until we provide a particular mechanism by which the fermion
field $\psi$ gets a mass, we have no definite prescription for fixing
$\alpha$. For example, in the five dimensional Kaluza-Klein theory the
term proportional to $\Gamma^{\hat\theta}\partial_\theta\Psi$ in
(\ref{sc}) can be interpreted as a mass term after a  $\alpha=\pi/2$ chiral rotation (\ref{chirrot}). The anomalous term transforms  as
\beq
P^\dagger
\l(i\gamma^0\sigma^{\mu\nu} \gamma^5\r)P=\gamma^0(\sin 2\alpha+
i\gamma^5\cos 2\alpha) \sigma^{\mu\nu}
\eeq
under the
chiral rotation (\ref{chirrot}).
Thus, in general, the anomalous
coupling (\ref{red_act}) is
\beq
\label{Action} {\cal
  S}_{int}= e \int d^4x F_{\mu\nu} \overline\psi(R_\chi \sin
2\alpha+iR_\chi \cos 2\alpha\gamma^5)\sigma^{\mu\nu}\psi,
\eeq
where $e$ is the electron charge
(we have adopted the usual normalization for the electromagnetic field).
The parameters $A$ and $B$ (\ref{anom1}) are respectively
\begin{eqnarray}
A&=&e R_\chi \sin{2\alpha} \\ \nn
B&=&e R_\chi \cos{2\alpha}.
\end{eqnarray}
As mentioned previously, this term gives a contribution to the
anomalous magnetic moment of the leptons as well as to their electric
dipole moments.

The contributions of this nonminimal coupling to the anomalous
magnetic moment of a fermion and to its electric dipole moment can
be deduced directly from (\ref{Action}). One gets

\begin{eqnarray}
   a^{\mbox{KK}}_\psi &=& R_\chi m_\psi \sin{2 \alpha}
\end{eqnarray}
for the anomalous magnetic moment of the fermion and
\begin{eqnarray}
   d^{\mbox{KK}}_\psi &=& \frac{e}{2} R_\chi \cos{2 \alpha}
\end{eqnarray}
for the electric dipole moment the fermion. The mass of the fermion is
denoted by $m_\psi$.

We shall first discuss the electric dipole moment.  The electric
dipole moment of the electron and of the muon have been measured very
precisely \cite{Groom:2000in}
\begin{eqnarray}
   d^{\ \mbox{exp}}_e &<& (0.18 \pm 0.12 \pm 0.10) \times 10^{-26}
   e-\mbox{cm}, \\
   d^{\ \mbox{exp}}_\mu &<& (3.7 \pm 3.4) \times 10^{-19} e-\mbox{cm}.
 \end{eqnarray}
 The same precision has not  been achieved for the $\tau$-lepton:
   \begin{eqnarray}
  d^{\ \mbox{exp}}_\tau &<& 3.1 \times 10^{-16} e-\mbox{cm}.
\end{eqnarray}
Using the experimental data, one gets
\begin{eqnarray} \label{res1}
R_e \cos{2 \alpha}<4.6
\times 10^{-14} \ \mbox{GeV}^{-1}
\end{eqnarray}
for the electron,
\begin{eqnarray}
  R_\mu \cos{2 \alpha}<
9.4 \times
10^{-6} \ \mbox{GeV}^{-1}
\end{eqnarray}
for the muon and
\begin{eqnarray}
  R_\tau \cos{2 \alpha}< 7.9 \times 10^{-3} \ \mbox{GeV}^{-1}
\end{eqnarray}
for the $\tau$-lepton. Note that $R_\chi$ (\ref{R}), $\chi=e, \mu,
\tau$, can take different values for different fermions. The best
limit is obtained for the electron electric dipole moment.

We now consider their anomalous magnetic moments.  Let us first
consider the electron. The value of the anomalous magnetic moment of
the electron predicted by the Standard Model is strongly dependent on
the value of the fine-structure constant \cite{Hughes:1999fp}.
Typically one gets
\begin{eqnarray} \label{expreselec}
  \Delta{a_e}=a_e(\mbox{exp})-a_e(\mbox{SM})
  &=& 34 (33.5) \times 10^{-12},
\end{eqnarray}
taking the fine-structure constant from the quantum Hall effect
measurement. Thus we get
\begin{eqnarray} \label{res4}
R_e \sin{2 \alpha} &<& \frac{\Delta{a_e}}{m_e}
= 6.7 \times 10^{-8} \ \mbox{GeV}^{-1}.
\end{eqnarray}

In the case of the muon, we can interpret the observed deviation
(\ref{expres}) as an effect of the nonminimal coupling. One obtains the
following constraint for the product $R_\mu \sin{2 \alpha}$
\begin{eqnarray} \label{res2}
R_\mu \sin{2 \alpha} &\le& \frac{\Delta{a_\mu}}{m_\mu}
= 4 \times 10^{-8} \ \mbox{GeV}^{-1}.
\end{eqnarray}
Besides the magnitude of the deviation from the Standard Model
prediction for the anomalous magnetic moment, its sign is of crucial
importance. This sign depends on the angle $\alpha$ (\ref{chirrot})
which is determined by the
fermion mass generating mechanism. Therefore, one gets a constraint on
model building $\alpha \in [0,\frac{\pi}{2}]$.  One cannot obtain
constraints from the anomalous magnetic moment of the $\tau$-lepton.

From the above estimates we see that the magnitude $R_\chi$
(\ref{red_act}) of the nonminimal coupling of fermions to the
electromagnetic field is less than or of the order of $ 10^{-21}$ cm.
Remember, that $R_\chi$ characterises the higher-dimensional profile
of a fermion. For example, in the models with large extra dimensions
it is an estimate of the size of the region where the fermions are
localized (``thickness'' of the brane).

We would like to point out that a term (\ref{anom1}) arises also for
neutrinos. Thus, neutrinos are expected to have a magnetic moment.
Using the available limit \cite{Groom:2000in} for the magnetic moment
of an electron-type neutrino, one gets
\begin{eqnarray} \label{res5}
R_\nu \sin{2 \alpha} &<& 3 \times 10^{-7} \ \mbox{GeV}^{-1}.
\end{eqnarray}
A similar constraint is obtained for the $\mu$-type neutrino. More
stringent limits can be obtained from astrophysical considerations
\cite{Raffelt:1999gv}.

It is important to note that in models with large extra dimensions,
there are extra contributions to the anomalous magnetic moment which
come from Kaluza-Klein excitations of bulk fields (e.g. bulk gravitons)
\cite{extra,Graesser:2000yg} which can result in effects of the same
order-of-magnitude as (\ref{expres}).

\section*{Acknowledgements}
We would like to thank H. Fritzsch and A. Hebecker for useful
discussions.  The work of A.N. was supported by SFB 375 der Deutsche
Forschungsgemeinschaft. X.C. is supported by the Deutsche
Forschungsgemeinschaft, DFG-No. FR 412/27-1.

\end{document}